\documentclass[final,5p,times]{elsarticle}

\usepackage{amssymb}
\usepackage{amsmath}
\usepackage[ruled,linesnumbered]{algorithm2e}
\usepackage{hyperref}
\usepackage{float}
\usepackage[switch]{lineno}
\usepackage{placeins}
\usepackage[parfill]{parskip}
\usepackage{multirow}
\usepackage{xcolor}
\journal{Results in Engineering}

\begin{document}
\begin{frontmatter}
\title{Text2Structure3D: Graph-Based Generative Modeling of Equilibrium Structures with Diffusion Transformers}

\author[inst1,mun_ds]{Lazlo Bleker\corref{cor1}}
\ead{lazlo.bleker@tum.de}
\author[inst2]{Zifeng Guo}
\author[inst3]{Kaleb E. Smith}
\author[inst4]{Kam-Ming Mark Tam}
\author[inst5]{Karla Saldaña Ochoa}
\author[inst1,mun_ds,inst6]{Pierluigi D'Acunto}
\affiliation[inst1]{organization={Technical University of Munich, School of Engineering and Design, Professorship of Structural Design},
            addressline={Arcisstraße 21}, 
            city={Munich},
            postcode={80333},
            country={Germany}}
\affiliation[mun_ds]{organization={Technical University of Munich, Munich Data Science Institute},
            addressline={Walther-von-Dyck-Straße 10}, 
            city={Garching},
            postcode={85748},
            country={Germany}}
\affiliation[inst2]{organization={Strabag AG},
            addressline={Albstadtweg 3}, 
            city={Stuttgart},
            postcode={70567},
            country={Germany}}
\affiliation[inst3]{organization={NVIDIA},
            city={Gainesville},
            country={USA}}
\affiliation[inst4]{organization={University of Hong Kong, Department of Architecture},
            addressline={4/F, Knowles Building, Pokfulam Road}, 
            city={Hong Kong SAR},
            country={China}}
\affiliation[inst5]{organization={University of Florida, School of Architecture, College of Design, Construction and Planning},
            city={Gainesville},
            country={USA}}
\affiliation[inst6]{organization={Technical University of Munich, Institute for Advanced Study},
            addressline={Lichtenbergstraße 2a}, 
            city={Garching},
            postcode={85748},
            country={Germany}}
\cortext[cor1]{Corresponding Author}

\begin{abstract}
This paper presents Text2Structure3D, a graph-based Machine Learning (ML) model that generates equilibrium structures from natural language prompts.
Text2Structure3D is designed to support new intuitive ways of design exploration and iteration in the conceptual structural design process.
The approach combines latent diffusion with a Variational Graph Auto-Encoder (VGAE) and graph transformers to generate structural graphs that are close to an equilibrium state.
Text2Structure3D integrates a residual force optimization post-processing step that ensures generated structures fully satisfy static equilibrium.
The model was trained and validated using a cross-typological dataset of funicular form-found and statically determinate bridge structures, paired with text descriptions that capture the formal and structural features of each bridge.
Results demonstrate that Text2Structure3D generates equilibrium structures with strong adherence to text-based specifications and greatly improves generalization capabilities compared to parametric model-based approaches.
Text2Structure3D represents an early step toward a general-purpose foundation model for structural design, enabling the integration of generative AI into conceptual design workflows.
\end{abstract}


\end{frontmatter}

\section{Introduction}
Load-bearing structures have one of the largest contributions to the total embodied carbon of buildings and infrastructure \cite{kaethner2012}, highlighting the importance of efficient structural designs.
Some of the largest structural design inefficiencies stem from decisions made during the conceptual design phase, where the global geometry is defined \cite{paulson1976, fang2023}.
Completely neglecting structural considerations in this critical design phase often leads to free-form geometries that work only with unnecessarily material-intensive supporting systems characterized by bending-dominant load-bearing behavior.
At the same time, design methods that do combine structural performance with global geometry, such as structural form-finding, typically require expert knowledge that may not always be available during the conceptual design stage.
Thus, developments aimed at making structure-informed design more accessible in the conceptual structural design phase can potentially have an outsized impact on the sustainability of the Architecture, Engineering and Construction (AEC) industry as a whole.

Generative Artificial Intelligence (GenAI) presents a unique opportunity for enhancing the conceptual structural design process and is rapidly transforming a range of other design domains, most prominently those dealing with images, video and natural language, but also areas such as 3D modeling and molecular design.
Within the AEC industry, GenAI is already being applied for general architectural design \cite{zhuang2025,khan2025}, automated floor plan generation \cite{nauata2020house,weber2022} and generative Building Information Modeling (BIM) \cite{du2025,liang2025}.
While there are also generative AI applications for structural design, they are generally focused on specific tasks, limiting their rate of adoption in structural design practice.

In this regard, general-purpose generative AI for structural design, especially models accessible through intuitive inputs such as text prompts, can enable a wider exploration of structural alternatives already in the conceptual design phase.
In particular, by generating structures that are informed by equilibrium and dominated by efficient axial force action, such models can help bridge the gap between intuitive design exploration and material-efficient structural design.

\subsection{Problem Statement}
Current generative AI approaches for structural design are often narrow in scope and typically confined to data representations dependent on problem-specific parametric models.
This makes such approaches time-intensive to adapt to new applications, requiring the retraining of models, creation of new datasets, or even adaptation of the model architecture.
General-purpose foundation models on the other hand represent a new paradigm in Machine Learning (ML) that aims to address these challenges by training a single model on large amounts of data that is flexible enough to adapt to a broad range of tasks.
However, existing foundation models such as Large Language Models (LLM) and image generation models are ill-suited for structural design applications.
Most foundation models operate on text and image modalities, which are poorly suited for representing structural systems compared to more flexible data structures such as graphs.
Moreover, the generative architectures underlying foundation models, such as transformers \cite{vaswani2017} and diffusion models \cite{sohl-dickstein2015, ho2020, song2021}, are general-purpose and typically lack explicit incorporation of domain-specific knowledge.
Most relevant to this work, diffusion-based generative models have not yet been combined with static equilibrium constraints fundamental to structural design.
Thus, developing generative methods that are both broadly applicable within structural design and aware of such relevant physical constraints remains a key challenge.

\subsection{Objectives and Scope}
\begin{figure}[h!]
    \centering
    \includegraphics[width=\linewidth]{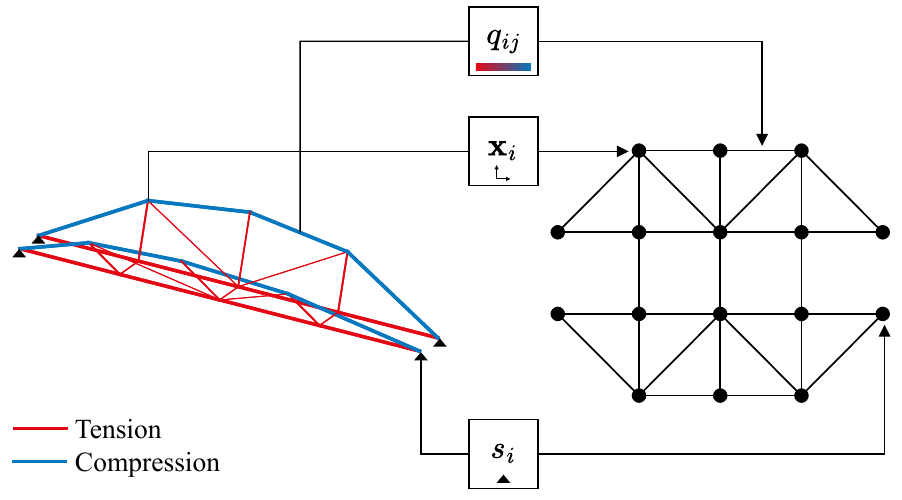}
    \caption{An equilibrium structure (left) represented as a graph (right). Every node $i$ of the graph is embedded with the coordinates ($\mathbf{x}_i$) and support condition ($s_i$) of the corresponding node in the equilibrium structure. Every edge $(i,j)$ of the graph contains the force density ($q_{ij}$) of the corresponding member in the equilibrium structure.} 
    \label{fig:graph_data}
\end{figure}

To address the challenges of applying generative AI in structural design we present \textit{Text2Structure3D}, a generative latent diffusion model \cite{rombach2022} for equilibrium structures conditioned on natural language.
Text2Structure3D operates on general graph-based data representations of discrete pin-jointed structures in line with the level of abstraction common in the conceptual design phase (Figure \ref{fig:graph_data}).
The scope of this study is restricted to form-found structures subject to a single dominant load case.
We specifically focus on bridge structures and assume a constant line load applied to the bridge deck.
We do not consider statically indeterminate structures or material parameters, and instead focus on satisfying static equilibrium as the most fundamental structural design constraint.

We condition the model on text embeddings to allow for the generation of equilibrium structures that can satisfy a variety of user-defined specifications.
In this way, Text2Structure3D is intended as a design support method for early-stage structural design. For architects, the natural language interface can enable intuitive exploration of structurally plausible forms before detailed engineering input is available. For structural engineers, the method can support rapid iteration during conceptual form-finding, especially for complex spatial structures where generating equilibrium geometries can otherwise be challenging and time-consuming.

\subsection{Contributions}
This study contributes to the field in the following three ways:
\begin{enumerate}
    \item By conditioning our model on text embeddings, Text2Structure3D enables new and intuitive natural language-based conceptual structural design processes. Structures adhering to a wide variety of design specifications can be generated, including text descriptions containing both qualitative and quantitative attributes, 
    \item We introduce a graph-based model architecture (Text2Structure3D) consisting of latent diffusion \cite{rombach2022}, a Variational Graph Auto-Encoder (VGAE) \cite{kipf2016}, and graph transformers \cite{dwivedi2020}, that can generate forms and forces of structures in static equilibrium. The combination of a VGAE with diffusion enables the generation of structures from a variable-sized and connected graph space that could not be generated using a VGAE alone.
    \item We jointly encode (continuous) edge force densities, (continuous) node coordinates and (categorical) support conditions into a shared node-level latent space that simplifies the latent diffusion process. The static equilibrium of reconstructions from this latent space is ensured by a residual force optimization post-processing step that combines the accuracy of geometry-based reconstructions with the physical validity of force-based reconstructions.
\end{enumerate}

We train and validate Text2Structure3D on a cross-typological synthetic dataset of combinations of bridge structures and text descriptions.
Together, Text2Structure3D represents a first step towards a general-purpose foundation model for the design of equilibrium structures.

\section{Related Work}
\label{sec:related_work}

\subsection{Generative Structural Design}
The term generative design has been used in architectural and structural design since well before the most recent generative AI revolution \cite{shea2005, krish2011}.
These early generative design approaches are often based on parametric models \cite{woodbury2010} or structural grammars \cite{shea1999}, encoding the structure as a (top-down) vector of design parameters, or a (bottom-up) sequence of design actions respectively.
Generative design typically involves the creation of designs that satisfy certain criteria or optimize a given objective. 
In this regard also form-finding \cite{veenendaal2012}, topology optimization \cite{bendsøe1988}, and layout optimization \cite{dorn1964}, can be considered as forms of generative design.
To satisfy design objectives, design spaces created with parametric models are commonly explored using metaheuristics such as evolutionary algorithms \cite{goldberg1989, hajela1995}.

More recently, generative AI approaches have been proposed as a powerful alternative for design space exploration.
Building on previous work for performance-based design exploration \cite{brown2019}, Danhaive and Mueller were one of the first to apply generative AI in structural design by training a Variational Auto-Encoder (VAE) \cite{kingma2013} to encode structural designs into a latent space \cite{danhaive2021}.
Bucher et al. extend this idea by conditioning the VAE on explicit performance parameters \cite{bucher2023}, and Balmer et al. additionally include Explainable AI (XAI) methods \cite{adadi2018, balmer2024}.
Saldana Ochoa et al. propose a structural form-finding-based generative design workflow using a discriminative ML model \cite{saldanaochoa2021}, and Guo et al. build on this approach with a Generative Adversarial Network (GAN) \cite{goodfellow2014} conditioned on single word text embeddings \cite{guo2022}.
Common to all these approaches is that they are based on parametric models that encode structural designs as vectors of model-specific design spaces.
New tasks involving parametric models other than the one for which the ML model was designed will produce incompatible vector data.
This presents a critical scalability limitation for parametric model-based approaches, invariably requiring retraining the model and adjusting its architecture for each new design task.

Moving beyond this limitation of vector data requires encoding structures with a more general data structure.
One alternative to parametric model-based approaches is to represent structures with higher rank tensors, such as images, instead.
Examples of structural design applications using generative ML models for image data include optimal layout design for shear walls \cite{liao2021} and steel frames \cite{fu2024}, and shell structure design \cite{mirra2021}.
Yet, image-based approaches still rely on a fixed parameterization scheme that does not accommodate design spaces involving systems of variable sizes and connectivities.
Consequently, they are ill-suited for modeling trans-topological design spaces.
Naive strategies that employ image-based methods for such systems often rely on fixed sampling techniques, which lack expressivity, lead to data loss, and are unable to provide fine-grained predictions at each element, as the sampling rate does not align neatly with the system's elements.
These limitations of Euclidean tensor-based data structures motivate the exploration of alternative more general structural design representations.

\subsection{Graph Machine Learning for Structural Design}
Unlike vectors or images, graphs represent a natural way of encoding discrete structures, especially those consisting of truss or beam elements which are ubiquitous in conceptual structural design \cite{tam2025}.
At the same time, Graph Neural Networks (GNN) \cite{kipf2016} allow ML models to make predictions on graphs, which has been explored for a number of graph-based structural design applications.
Chang and Cheng use GNNs to accelerate cross-section optimization \cite{chang2020}, and Whalen and Mueller train a GNN as a surrogate model for truss Finite Element Method (FEM) \cite{whalen2021}.
Hayashi et al. combine GNNs with Reinforcement Learning (RL) for cross-section \cite{hayashi2022} and assembly sequence optimization \cite{hayashi2022assembly}, Zhao et al. use GNNs for automated shear wall layout design \cite{zhao2023}, and Xia et al. extend this direction with a GNN-based co-design method for jointly generating shear wall and beam layouts \cite{xia2025}.
Li et al. combine expert-informed GNNs with metaheuristic optimization for the generative design of modular construction structures \cite{li2025}.
Bleker et al. apply GNNs to structural form-finding using the Combinatorial Equilibrium Modeling (CEM) \cite{ohlbrock2020}, first with a discriminative model \cite{bleker2022} and later extend it with a generative model \cite{bleker2024}.

Graph-based models also naturally allow for integration with knowledge and constraints from statics and geometry.
Tam et al. introduce a residual force loss that pushes predictions to equilibrium states \cite{tam2022trans,tam2026} and Bleker et al. apply an Equivariant Graph Neural Network (EGNN) \cite{satorras2021} to ensure predictions are equivariant to Euclidean transformations \cite{bleker2025generalized}.

Despite the powerful representation provided by graphs, generative models in structural design have up until now been more commonly based on tensor representations, such as vectors and images.
Liao et al. give the high level of complexity and sophistication of generative graph models as one of the reasons \cite{liao2024}.
However, recent developments in generative models for graphs have the potential of making graph models more suitable for generative structural design applications.

\subsection{Diffusion Models}
Diffusion models are a type of generative AI based on a gradual denoising process to obtain samples from a data distribution \cite{sohl-dickstein2015, ho2020, song2021}. 
Stemming from computer vision, diffusion models have been popularized by applications such as DALL-E 2 by OpenAI \cite{ramesh2022} and the open-source Stable Diffusion \cite{rombach2022}, and have since then been extended to other modalities, such as audio \cite{kong2021}, text \cite{li2022}, and, most relevant to this work, graphs \cite{vignac2023}.
The most prominent domains of graph diffusion are for molecular generation, such as material and drug design, where they have become the state of the art \cite{hoogeboom2022, wu2022}.
A common approach is to apply diffusion at the node-level feature space of graphs.
However, this presents the challenge that features often consist of a mix of continuous and discrete variables, requiring a sophisticated and complex diffusion process \cite{anand2022}.
To overcome this limitation, Xu et al. propose applying latent diffusion on graphs by encoding node features into a node-level latent space \cite{xu2023}.
In this way, diffusion on a complex categorical-continuous product manifold is avoided in favor of a purely continuous latent space.
This approach simplifies the modeling process and at the same time allows graphs of various types to be modeled by a single ML model \cite{joshi2025}.

\begin{figure*}[h]
    \centering
    \includegraphics[width=\linewidth]{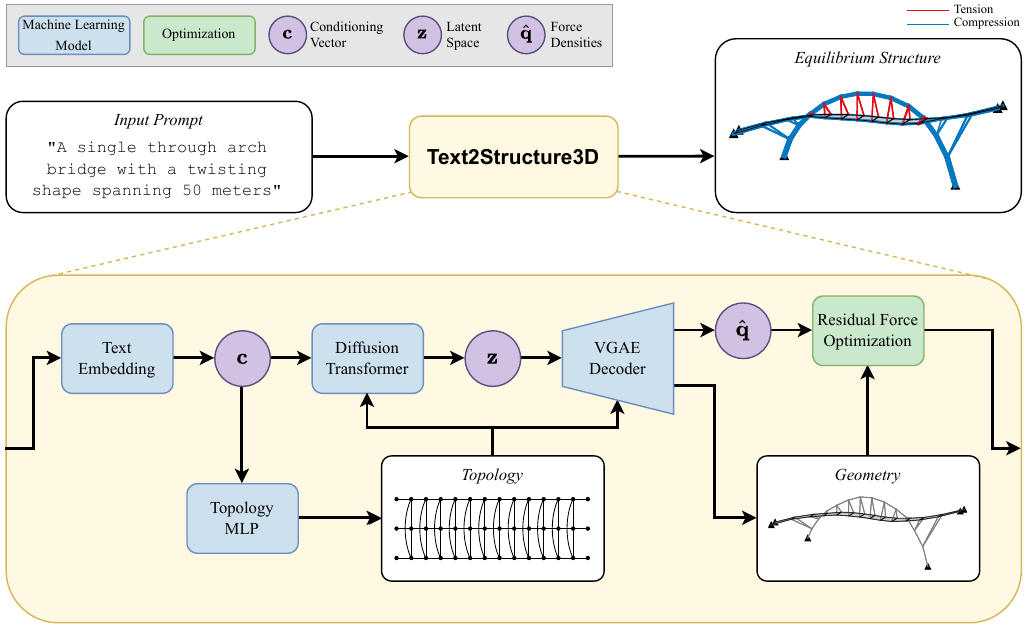}
    \caption{Text2Structure3D can generate an equilibrium structure adhering to an input text prompt. A text embedding model creates a conditioning vector ($\mathbf{c}$) that is input to a diffusion transformer and a topology Multilayer Perceptron (MLP). The diffusion transformer denoises a sample in the latent space ($\mathbf{z}$), conditioned on $\mathbf{c}$ and a topology generated by the topology MLP. A VGAE Decoder \cite{kipf2016}, also conditioned by the topology, creates a geometry and set of force densities ($\hat{\mathbf{q}}$) from the latent. The geometry and $\hat{\mathbf{q}}$ are optimized together to obtain a final sample of a structure in equilibrium. Line thickness corresponds to force magnitudes in tension (red) and compression (blue).}
    \label{fig:text2struct_overview}
\end{figure*}

Similar to molecular data, graph-based structural data also consists of both continuous node features (coordinates) as well as categorical features (support conditions).
Moreover, unlike for molecules in which edge features can be modeled implicitly, structural data requires explicitly defined edge features, such as forces or force densities.
In this work, we apply latent diffusion to graph-based equilibrium structures, encoding both categorical and continuous features, as well as node and edge features into one combined node-level latent space.
\section{Latent Diffusion for Equilibrium Structures}
\label{sec:methods}
Text2Structure3D is a graph-based latent diffusion model that can generate equilibrium structures from text prompts (Figure \ref{fig:text2struct_overview}).
During inference, a conditioning vector ($\mathbf{c}$) is created by a pre-trained text embedding model.
A Diffusion Transformer (DiT) conditioned on $\mathbf{c}$ then generates a point in the latent space ($\mathbf{z}$), which is decoded into a structural geometry and set of force densities ($\hat{\mathbf{q}}$) by a VGAE decoder.
Both the DiT and VGAE receive an additional topology input produced by a Multilayer Perceptron (MLP) conditioned on $\mathbf{c}$ in a parallel stream.
Finally, a residual force optimization post-processes the geometry and force densities ($\hat{\mathbf{q}}$) to obtain the final equilibrium structure.

The individual models of Text2Structure3D are trained in a reverse order relative to the order used during inference.
First, we train the VGAE to encode equilibrium structures into a latent space (Section \ref{sec:vgae}), and to be able to reconstruct these with the help of residual force optimization to preserve equilibrium (Section \ref{sec:optim}). Then, we freeze the VGAE weights and use its encoder by training the DiT to generate samples from the distribution of the VGAE latent space (Section \ref{sec:dit}). In Section \ref{sec:top} we outline the topology MLP and finally in Section \ref{sec:data} we explain the creation of the synthetic dataset on which we train Text2Structure3D.

\subsection{Variational Graph Auto-Encoder}
\label{sec:vgae}
The first step of Text2Structure3D involves training a Variational Graph Auto-Encoder (VGAE) \cite{kipf2016} to encode equilibrium structures into a latent space.
Equilibrium structures are modeled as graphs with a mix of node- and edge-level, as well as continuous and categorical features:
\begin{equation}
    \begin{aligned}
    \text{Force density}    \quad\mathbf{q} &= \{\, q_{ij} \mid (i,j)\in\mathcal{E} \,\} &&\in \mathbb{R}^{E},\\
    \text{Coordinates}     \quad\mathbf{x} &= \{\, \mathbf{x}_i \mid i=1,\dots,N \,\}             &&\in \mathbb{R}^{N\times 3},\\
    \text{Support condition} \quad\mathbf{s} &= \{\, s_i \mid i=1,\dots,N \,\}            &&\in \mathbb{Z}^{N}.
    \end{aligned}
\end{equation}

Here, $N$ is the number of nodes and $\mathcal{E}$ the set of edges ($|\mathcal{E}|=E$) that defines the structure's topology. 
Each edge $(i,j)\in\mathcal{E}$ is assigned a force density $q_{ij}$ (kN/m), while each node $i$ is associated with coordinates $\mathbf{x}_i$ (m) and a categorical support condition $s_i\in\{0,1\}$,  where $0$ denotes a node as free and $1$ as fixed.
Here, the force density of an edge is defined as the axial force in the corresponding member ($f_{ij}$) divided by its length ($l_{ij}$):
\begin{equation}
q_{ij}=\frac{f_{ij}}{l_{ij}}.
\end{equation}
We use the edge force density for its linear relation with $\mathbf{x}$ \cite{schek1974}, though an edge force definition results in an equivalent representation and could also be used.
It should be noted we do not explicitly model the external load, and instead derive it from the geometry based on assumptions of the training dataset (see Section \ref{sec:data}).

\begin{figure}[h]
    \centering
    \includegraphics[width=\linewidth]{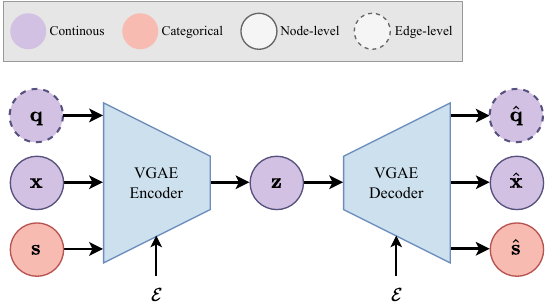}
    \caption{A Variational Graph Auto-Encoder (VGAE) \cite{kipf2016} learns a node-level latent space by jointly reconstructing both node-level and edge-level, as well as continuous and categorical features. The features of our equilibrium data consist of the edge force density ($\mathbf{q}$), the node coordinates ($\mathbf{x}$) and the node support condition ($\mathbf{s}$).}
    \label{fig:text2struct_vae}
\end{figure}

The VGAE encoder takes these multi-modal features as inputs and maps them into a shared node-level latent space $\mathbf{z}\in\mathbb{R}^{N\times d_z}$, see Figure~\ref{fig:text2struct_vae}.
The decoder then takes $\mathbf{z}$ and jointly reconstructs the features $\{\hat{\mathbf{q}}, \hat{\mathbf{x}}, \hat{\mathbf{s}}\}$.

We train the VGAE with a reconstruction loss, using a Mean Squared Error (MSE) for $\hat{\mathbf{q}}$ and $\hat{\mathbf{x}}$, and a Binary Cross Entropy (BCE) for $\hat{\mathbf{s}}$:
\begin{align}
\mathcal{L}_q &= \tfrac{1}{E}\sum_{(i,j)\in\mathcal{E}} (\hat q_{ij}-q_{ij})^{2},\\[6pt]
\mathcal{L}_x &= \tfrac{1}{N}\sum_{i=1}^{N}\|\hat{\mathbf{x}}_{i}-\mathbf{x}_{i}\|_2^{2},\\[6pt]
\mathcal{L}_s &= \tfrac{1}{N}\sum_{i=1}^{N}\Big(-s_i \log \sigma(\hat{s}_i) \notag \\[-2pt]
              &\qquad\qquad - (1-s_i)\log(1-\sigma(\hat{s}_i))\Big).
\end{align}

Furthermore, we apply a KL divergence loss to weakly regularize the latent space according to
\begin{equation}
\mathcal{L}_{\text{KL}} = D_{\mathrm{KL}}\!\Big(\mathcal{N}(\boldsymbol{\mu}_z,\boldsymbol{\sigma}_z)\,\Big\|\,\mathcal{N}(\mathbf{0},\mathbf{I})\Big).
\end{equation}

The total loss is then described as the following weighted sum:
\begin{equation}
\mathcal{L} = \lambda_q \mathcal{L}_q
            + \lambda_x \mathcal{L}_x
            + \lambda_s \mathcal{L}_s
            + \beta \mathcal{L}_{\text{KL}}.
\end{equation}

Because the distribution of encoded structures is modeled by the subsequent diffusion model, the Gaussian prior of the VGAE is not used as the direct sampling distribution during inference. Instead, following Rombach et al. \cite{rombach2022}, the KL divergence term is retained as a weak regularizer to prevent an arbitrarily high-variance latent space, which can hinder diffusion-based generation. We therefore prioritize accurate reconstructions and set the KL divergence weight to a low value. We found $\beta = 0.01$ to produce good results, combined with $\lambda_q = \lambda_x = \lambda_s = 1.0$.
For the architecture of both the encoder and decoder, we use a graph transformer following the design of Rampá\v{s}ek et al. \cite{rampasek2022}.
For additional hyperparameter details, we refer to \ref{sec:apdx_implementation_details}.

\subsection{Equilibrium-based Reconstruction}
\label{sec:optim}
The reconstructed force densities $\hat{\mathbf{q}}$ will in general not perfectly match the ground truth $\mathbf{q}$, and similarly, $\hat{\mathbf{x}}$ will not equal $\mathbf{x}$.
While a limited reconstruction error in both $\hat{\mathbf{q}}$ and $\hat{\mathbf{x}}$ is admissible, uncoordinated errors will lead to structures that are not in equilibrium.
To determine if a structure is in equilibrium we can calculate its residual force according to
\begin{equation}
    \mathbf{r}_i = \mathbf{p}_i 
    + \sum_{j \in \mathcal{N}(i)} q_{ij}\,(\mathbf{x}_j - \mathbf{x}_i),
\end{equation}

where $\mathbf{r}_i$ and $\mathbf{p}_i$ are the residual force and applied load at node $i$, respectively.
In order for a structure to be in equilibrium, the magnitude of the residual force must vanish at every free node, i.e.
\begin{equation}
\begin{aligned}
\|\mathbf{r}_i\|_2 = 0, \quad \forall i \in \mathcal{S}_{\mathrm{free}}, \quad \mathcal{S}_{\mathrm{free}} := \{\, i \mid \hat{s}_i < 0 \,\}.
\end{aligned}
\end{equation}

To prevent non-equilibrium structures, we apply a post-processing optimization step. Specifically, we perform a combined optimization of $\tilde{\mathbf{x}}$ and $\tilde{\mathbf{q}}$ and minimize the squared residual force magnitudes of free nodes:
\begin{equation}
\begin{aligned}
\min_{\tilde{\mathbf{x}},\,\tilde{\mathbf{q}}} \quad & 
\sum_{i \in \mathcal{S}_{\mathrm{free}}} 
\left\| \, \mathbf{p}_i 
+ \sum_{j \in \mathcal{N}(i)} \tilde q_{ij}\,(\tilde{\mathbf{x}}_j - \tilde{\mathbf{x}}_i) \, \right\|_2^2 \\[2ex]
\text{s.t.} \quad & 
\hat{\mathbf{x}}_i - \Delta_x \;\leq\; \tilde{\mathbf{x}}_i \;\leq\; \hat{\mathbf{x}}_i + \Delta_x,
\quad \forall i, \\[1ex]
& \hat{q}_{ij} - \Delta_q \;\leq\; \tilde{q}_{ij} \;\leq\; \hat{q}_{ij} + \Delta_q,
\quad \forall (i,j)\in \mathcal{E}.
\end{aligned}
\end{equation}

We initialize $\tilde{\mathbf{x}}$ and $\tilde{\mathbf{q}}$ as the decoder outputs $\hat{\mathbf{x}}$ and $\hat{\mathbf{q}}$, and apply bounds around this initialization defined by $\Delta_x$ and $\Delta_q$. We find that typically only minor adjustments to both $\hat{\mathbf{x}}$ and $\hat{\mathbf{q}}$ are needed to obtain structures that are in equilibrium within a given tolerance and set $\Delta_x = 2$ m and $\Delta_q = 1$ kN/m to prevent the optimization from straying too far from the decoder prediction. Optionally, any small residual still remaining after optimization can be eliminated by applying a final form-finding step with the Force Density Method (FDM) \cite{schek1974}.

\begin{figure*}[h]
    \centering
    \includegraphics[width=\linewidth]{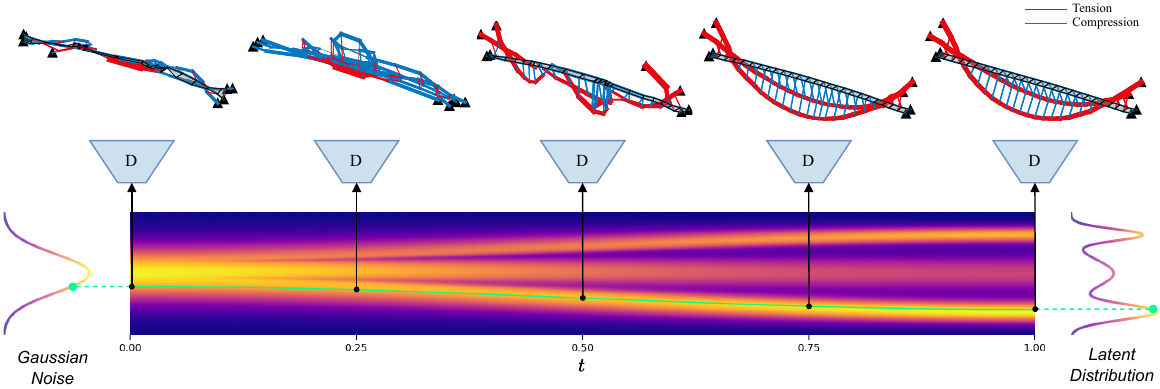}
    \caption{A random sample from a Gaussian distribution is iteratively denoised to a point in the latent space distribution. The VGAE decoder can decode denoised and intermediate samples into structures. Only structures decoded at the end of the diffusion process (right) will be close to an equilibrium state. Line thickness corresponds to force magnitudes in tension (red) and compression (blue).}
    \label{fig:text2struct_flowmatch}
\end{figure*}
\subsection{Diffusion Transformer}
\label{sec:dit}
After training the VGAE, we freeze its weights and treat its encoder and decoder as mappings between equilibrium structures and latents $\mathbf{z}$. Using this fixed latent space, we can model its distribution using a Gaussian diffusion framework, allowing pure noise to be mapped to the target latent distribution (Figure \ref{fig:text2struct_flowmatch}). Specifically, we use a flow-matching formulation that models the transformation from the Gaussian prior to the latent distribution as a time-dependent vector field \cite{lipman2023}. In this way, the diffusion process spans the gap between a structure sampled from the gaussian prior using only the VGAE decoder (Figure \ref{fig:text2struct_flowmatch}, left), to a structure sampled from an approximation of the actual VGAE latent space (Figure \ref{fig:text2struct_flowmatch}, right).

Given an equilibrium structure, we encode it to a target latent $\mathbf{z}^{(1)}$ at time $t=1$. We sample an initial noise latent $\mathbf{z}^{(0)}$ from a standard normal, and construct a linearly interpolated latent at a random time $t\sim\mathcal{U}(0 + \epsilon,1 - \epsilon)$ via
\begin{equation}
\mathbf{z}^{(t)} \;=\; (1-t)\,\mathbf{z}^{(0)} + t\,\mathbf{z}^{(1)},
\end{equation}

where we set $\epsilon = 0.01$ for numerical stability. Along this straight-line path, the ground-truth conditional vector field is
\begin{equation}
u_t\!\left(\mathbf{z}^{(t)} \mid \mathbf{z}^{(1)}\right) \;=\; \frac{\mathbf{z}^{(1)} - \mathbf{z}^{(t)}}{\,1-t\,}.
\end{equation}

The central idea behind flow-matching is to train a neural network to approximate this vector field conditional on $t$. During inference, starting with a noise sample we can then integrate along the approximate vector field to arrive at a denoised latent, as depicted by the green line in Figure \ref{fig:text2struct_flowmatch}.

To improve training stability, we follow Yim et al. \cite{yim2023} and reparameterize the flow-matching loss to predict the denoised latent according to
\begin{equation}
    \mathcal{L}_{\text{fm}} = \frac{1}{N} \sum_{i=1}^N 
    \left\|
        \frac{\mathbf{z}^{(1)}_i - \mathbf{z}^{(t)}_i}{1 - t}
        - \frac{\hat{\mathbf{z}}^{(1)}_i - \mathbf{z}^{(t)}_i}{1 - t}
    \right\|_2^2,
    \label{eq:fm_loss1}
\end{equation}

which simplifies to
\begin{equation}
    \mathcal{L}_{\text{fm}} = \frac{1}{(1 - t)^2} \cdot
    \frac{1}{N} \sum_{i=1}^N
    \left\| \mathbf{z}^{(1)}_i - \hat{\mathbf{z}}^{(1)}_i \right\|_2^2,
    \label{eq:fm_loss2}
\end{equation}

in which we clip $t$ to 0.9 for numerical stability.

For the neural network we use a Diffusion Transformer (DiT) \cite{peebles2023}.
To guide the diffusion process towards structures that match a given text prompt, we obtain a conditioning vector $\mathbf{c}$ from a pre-trained CLIP model \cite{radford2021}.
We use the OpenCLIP implementation \cite{ilharco2021} with the ViT-g/14 checkpoint trained on LAION-2B \cite{schuhmann2022}.
Conditioning on $\mathbf{c}$ and $t$ is injected into the DiT via adaptive layer normalization with zero-initialization.

We train the DiT with $\mathcal{L}_{\mathrm{fm}}$ in Equation~\eqref{eq:fm_loss2}, using the frozen VGAE decoder to obtain the target $\mathbf{z}_i^{(1)}$.
We randomly drop $\mathbf{c}$ in favor of a learned null embedding $\varphi$ 10\% of the time during training, allowing us to apply Classifier-Free Guidance (CFG) \cite{ho2022} during inference, resulting in improved prompt-adherence. Further details and hyperparameters of the DiT can be found in \ref{sec:apdx_implementation_details}.

At inference, we first sample $\mathbf{z}^{(0)}\sim\mathcal{N}(0,I)$ and integrate the learned vector field forward in time with $T$ steps of size $\Delta t=1/T$, where we set $T = 100$.
Using CFG, we evaluate both a conditional and an unconditional DiT pass and combine them to amplify the direction associated with the text embedding:
\begin{equation}
\begin{aligned}
\hat{\mathbf{z}}^{(1)}_{\text{cond}} &= \operatorname{DiT}\big(\mathbf{z}^{(t)},\,t,\,\mathbf{c},\,\mathcal{E}\big), \\
\hat{\mathbf{z}}^{(1)}_{\text{uncond}} &= \operatorname{DiT}\big(\mathbf{z}^{(t)},\,t,\,\varphi,\,\mathcal{E}\big), \\
\hat{\mathbf{z}}^{(1)} &= (1-\gamma)\,\hat{\mathbf{z}}^{(1)}_{\text{uncond}} \;+\; \gamma\,\hat{\mathbf{z}}^{(1)}_{\text{cond}},
\end{aligned}
\end{equation}

in which $\gamma$ is the guidance scale, which we set to 5.0. We apply a simple Euler integration scheme according to the pseudocode in Algorithm \ref{alg:sampling}. Finally, after $T$ steps we decode the denoised latent to an equilibrium structure with the frozen VGAE decoder, and apply the residual force post-optimization from Section~\ref{sec:optim}.

\begin{algorithm}[h]
\caption{Latent Diffusion Euler Sampling}
\label{alg:sampling}
\KwIn{Text embedding $\mathbf{c}$, number of steps $T$,\\guidance scale $\gamma$, edge set $\mathcal{E}$}
\KwOut{Equilibrium structure $(\tilde{\mathbf{q}},\tilde{\mathbf{x}},\hat{\mathbf{s}})$}

Sample initial noise: $\mathbf{z}^{(0)} \sim \mathcal{N}(\mathbf{0},\mathbf{I})$\;
Set step size $\Delta t = 1/T$\;

\For{$t \in \text{linspace}(0,1,T)$}{
    $\hat{\mathbf{z}}^{(1)}_{\text{cond}} = \operatorname{DiT}\big(\mathbf{z}^{(t)},\,t,\,\mathbf{c},\,\mathcal{E}\big)$\;
    $\hat{\mathbf{z}}^{(1)}_{\text{uncond}} = \operatorname{DiT}\big(\mathbf{z}^{(t)},\,t,\,\varphi,\,\mathcal{E}\big)$\;
    $\hat{\mathbf{z}}^{(1)} = (1-\gamma)\,\hat{\mathbf{z}}^{(1)}_{\text{uncond}} \;+\; \gamma\,\hat{\mathbf{z}}^{(1)}_{\text{cond}}$\;
    $\mathbf{z}^{(t+\Delta t)} = \mathbf{z}^{(t)} + \Delta t \,\dfrac{\hat{\mathbf{z}}-\mathbf{z}^{(t)}}{1-t}$\;
}
Decode: $(\hat{\mathbf{q}},\hat{\mathbf{x}},\hat{\mathbf{s}}) = \operatorname{D}(\mathbf{z}^{(1)},\,\mathcal{E})$\;
Residual force optimization: $(\tilde{\mathbf{q}},\tilde{\mathbf{x}}) = \operatorname{Optim}(\hat{\mathbf{q}},\hat{\mathbf{x}},\hat{\mathbf{s}})$\;
\end{algorithm}

\subsection{Topology Model}
\label{sec:top}
The final neural network of Text2Structure3D is a multilayer perceptron (MLP) that predicts a topology of the structure given the text embedding $\mathbf{c}$.
During inference, the MLP is the first model to be run following the text embedding model, so that the DiT and VGAE can use the predicted topology for their positional encodings and message passing layers (\ref{sec:apdx_implementation_details}).

Large groups of structures can all share the same topology through variation in their geometry and internal forces, or even their typologies, as is the case for arch and suspension bridges.
We make use of this fact to efficiently parameterize all the topologies in our dataset using a limited number of parameters.
This allows us to use a simple MLP to model the distribution of these topology parameters from $\mathbf{c}$.
For the model trained on the presented dataset the primary topology parameters are the number of bays, presence of diagonals, and single vs double arch/cable/chord configurations.
It should be noted that while this model design makes the topology MLP efficient, it does rely on a high-level parameterization of the topology, complicating generalization to more topologically complex datasets.

We train the MLP using a Cross Entropy (CE) loss, which converges to the topology parameter probability distribution conditional on $\mathbf{c}$.
Then, during inference, we can sample from this distribution to construct the predicted topology. For additional architecture and hyperparameter details of the MLP we also refer to \ref{sec:apdx_implementation_details}.

\begin{figure}[h!]
    \centering
    \includegraphics[width=\linewidth]{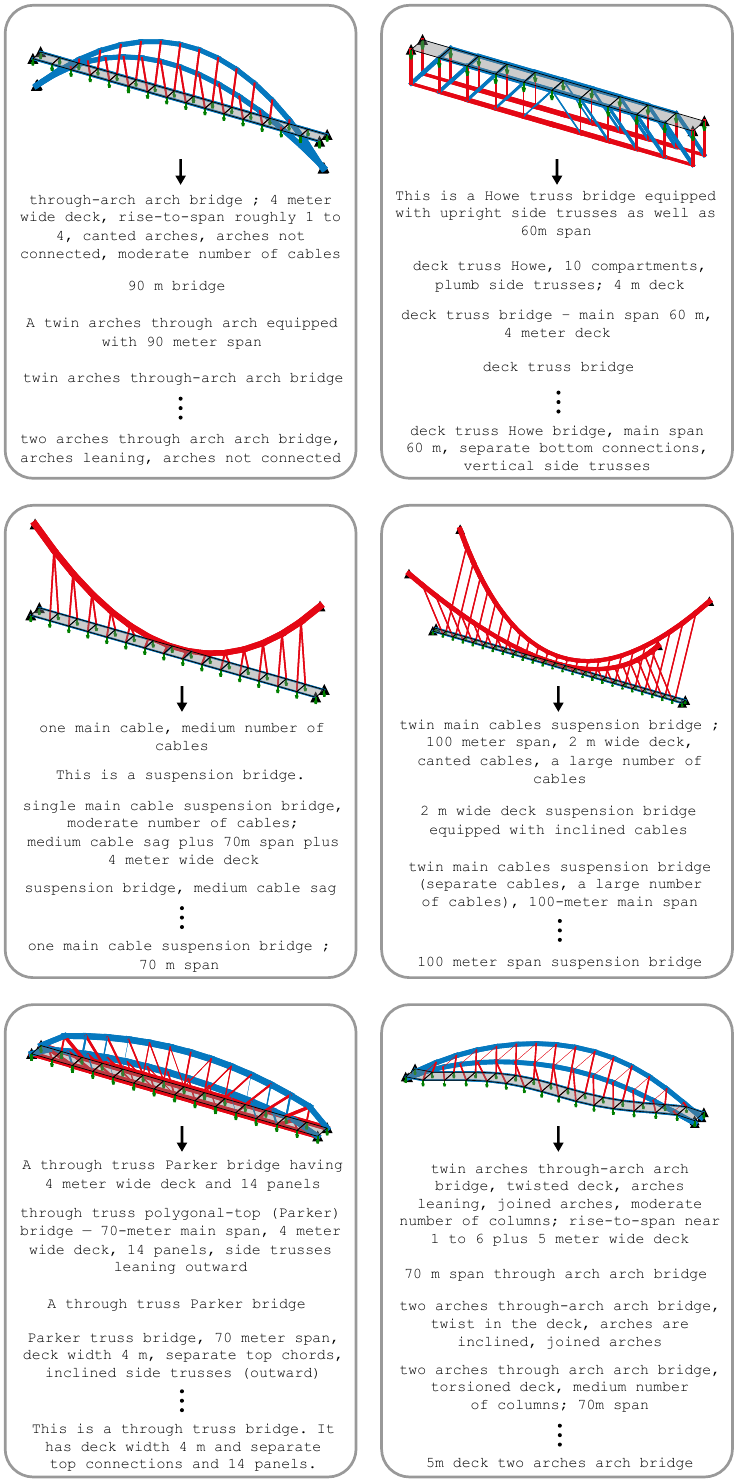}
    \caption{Samples from the cross-typological synthetic training dataset used to train Text2Structure3D. For every equilibrium structure we generate a number of text descriptions of various level of detail. Line thickness corresponds to force magnitudes in tension (red) and compression (blue).}
    \label{fig:data_array}
\end{figure}

\subsection{Synthetic Dataset}
\label{sec:data}
\begin{figure*}[h]
    \centering
    \includegraphics[width=\linewidth]{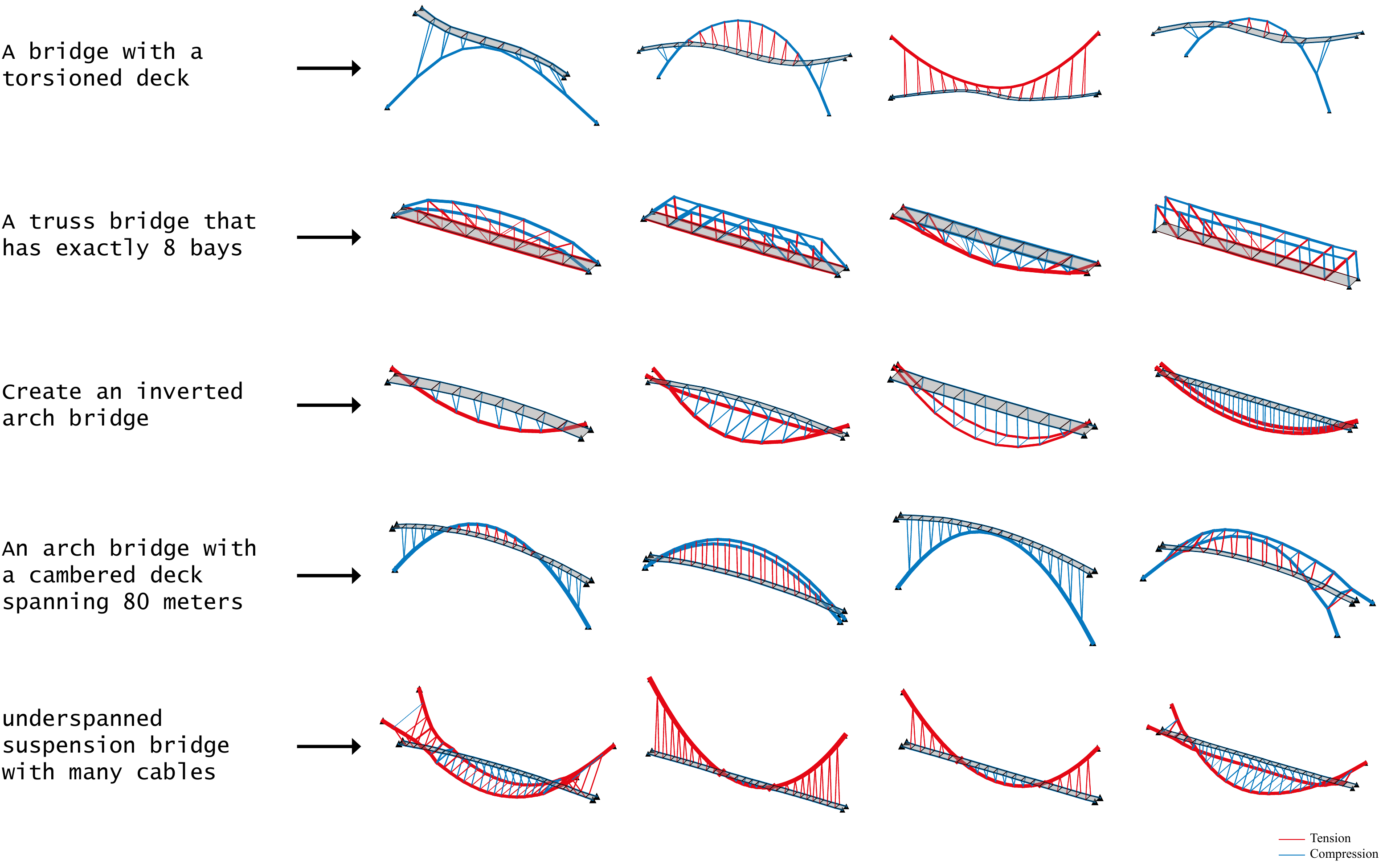}
    \caption{Equilibrium structures generated by Text2Structure3D using various text prompts. Text2Structure3D can generate diverse structures while maintaining adherence to the prompt specifications. Line thickness corresponds to force magnitudes in tension (red) and compression (blue).}
    \label{fig:results_main}
\end{figure*}
Due to the limited availability of publicly accessible datasets containing equilibrium structures across diverse typologies, we develop a cross-typological synthetic dataset of equilibrium structures and corresponding text labels to train Text2Structure3D (Figure \ref{fig:data_array}). The developed synthetic dataset is based on preliminary work \cite{bleker2026}, and consists of a combination of funicular arch and suspension bridge structures and statically determinate truss systems.

Funicular structures are form-found with the Combinatorial Equilibrium Modeling (CEM) \cite{ohlbrock2020}.
By varying the CEM input parameters we can obtain arch and suspension bridge structures with various topologies, geometries and force configurations.
Some of the parameters we set include the span, deck width and number of cables or columns along the deck.
Bridges either have a single or double arch/cable configuration, for arch and suspension structures respectively.
In the case of a double arch/cable configuration, these can be separate or interconnected by additional horizontal members.
The deck can be twisted in the horizontal plane and/or cambered vertically and we allow various support configurations that have the arch(es) or main cable(s) above, below, or intersect the deck.
During data generation, we sample these form-finding parameters from relatively large bounds so as to maximize the variety of structural configurations in the dataset.
To prevent any unrealistic or otherwise undesired structures, we filter out any structures that exceed a specified bounding box, exceed a maximum force density threshold or bend too sharply in the case of twisted bridges.

Truss structures are generated using a parametric model that is purely based on geometric rules.
We include Pratt, Howe and Parker truss configurations and allow the trusses to be above (through truss), or below (deck truss) the deck.
Similar to funicular bridges we vary the span, deck width and number of bays.
The trusses on both sides of the deck can be either vertical, or inclined outwards or inwards.
Furthermore, for deck trusses we include a triangular truss configuration such that trusses on either side are connected, sharing the same bottom chord. 
Finally, to determine the internal member forces we assemble and solve for the global equilibrium matrix.

We apply a canonical orientation to all structures, setting the midspan at the height of the deck to the origin and aligning the deck along the x-axis. We apply a constant line load of 5 kN/m to each bridge, such that during inference it is possible to derive the load from the span of the generated geometry.

To generate text labels, we use a randomized template that combines a list of attributes of each bridge in a description.
We include both quantitative attributes, such as span length, deck width and the number of bays, as well as qualitative attributes, such as typology or whether the bridge is twisted.
For some attributes we include both quantitative and qualitative descriptions.
For example, the rise of an arch can be described as an exact ratio (e.g. 1:5), or qualitatively (e.g. shallow).
We generate text labels with a broad range of levels of detail, including descriptions that contain all known attributes of the structure as well as descriptions with only a subset or even just a single one (e.g. a 60 meter span bridge).
In this way, we facilitate both use cases in which Text2Structure3D could be prompted with a very detailed description, or a shorter simple request.

We generate 20,000 funicular structures and 10,000 truss structures and combine these into a single dataset of 30,000 structures (Table \ref{tab:dataset}).
For each structure we generate 10 unique text labels such that our dataset contains 300,000 structure-text pairs.
Finally, we split our data into an 80\% training set (24,000) and set aside 20\% (6,000) for testing.

\begin{table}[h]
\centering
\resizebox{\columnwidth}{!}{
\begin{tabular}{lccc}
\hline
Structure Type & \# Structures & Labels/Structure & \# Pairs \\
\hline
Funicular & 20,000 & 10 & 200,000 \\
Truss     & 10,000 & 10 & 100,000 \\
\hline
Total     & 30,000 & -- & 300,000 \\
\hline
\end{tabular}}
\caption{Dataset composition of funicular and truss structures with corresponding text labels.}
\label{tab:dataset}
\end{table}
\section{Results}
\label{sec:results}

After training, Text2Structure3D is able to generate a structure of a bridge in equilibrium solely from an input text prompt (Figure \ref{fig:results_main}).
As a visual validation, we find generated structures to be geometrically similar to the training dataset of bridge structures and that prompt specifications are satisfied. Generated structures are consistently of the requested topology and qualitative attributes, such as \textit{a torsioned deck}, are satisfied as well. To evaluate the results quantitatively we study the reconstruction and sampling performance (Section \ref{sec:text2structurereconstruction}) and the quantitative prompt adherence (Section \ref{sec:text2structurerepromptadh}) in more detail. 

\subsection{Reconstruction and Sampling}
\label{sec:text2structurereconstruction}
We evaluate reconstructions in two ways, by assessing if they are in equilibrium, and according to their geometric reconstruction error. Reconstructed structures should be in equilibrium, which we measure by the residual force error $\epsilon_{\mathrm{residual}}$, calculated as the average residual force magnitude among free nodes:
\begin{equation}
\epsilon_{\mathrm{residual}} 
= \frac{1}{|\mathcal{S}_{\mathrm{free}}|} \sum_{i \in \mathcal{S}_{\mathrm{free}}} \| r_i \|_2 ,
\end{equation}

where $\epsilon_{\mathrm{residual}} = 0$ corresponds to a structure in equilibrium.

\begin{figure}[h]
    \centering
    \includegraphics[width=\linewidth]{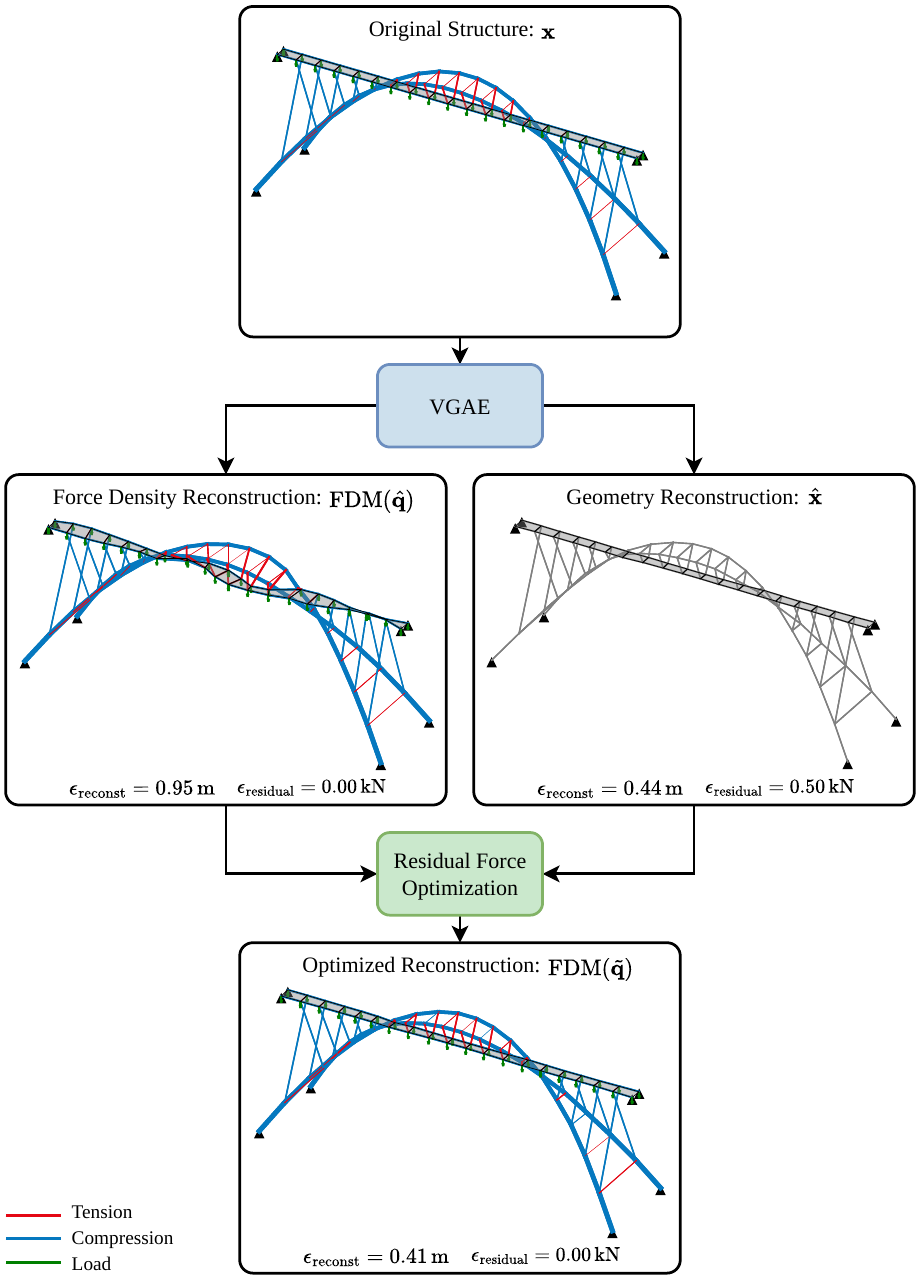}
    \caption{Variational Graph Auto-encoder (VGAE) reconstructions before and after residual force optimization. Force density-based reconstructions are guaranteed to be in equilibrium ($\epsilon_{\mathrm{residual}} = 0$), while geometry-based reconstructions have a lower reconstruction error ($\epsilon_{\mathrm{reconst}}$). We obtain a structure that combines the advantages of both by applying a residual force optimization post-processing step. Line thickness corresponds to force magnitudes in tension (red) and compression (blue).}
    \label{fig:results_reconstruct}
\end{figure}

The simplest way to obtain a reconstructed structure in equilibrium is by applying the Force Density Method (FDM) directly to the predicted $\hat{\mathbf{q}}$ produced by the VGAE (Figure \ref{fig:results_reconstruct}, left). However, the geometry produced by the FDM is very sensitive to small perturbations of $\hat{\mathbf{q}}$, which can result in asymmetric structures that are visually dissimilar to the original. We measure the quality of the reconstructed geometry using a reconstruction error $\epsilon_{\mathrm{reconst}}$, defined as the average nodal distance to the original structure:
\begin{equation}
\epsilon_{\mathrm{reconst}} 
= \frac{1}{N} \sum_{i=1}^{N} \| \hat{\mathbf{x}}_i - \mathbf{x}_i \|_2 .
\label{eq:epsilon_reconst}
\end{equation}

\begin{figure}[h]
    \centering
    \includegraphics[width=\linewidth]{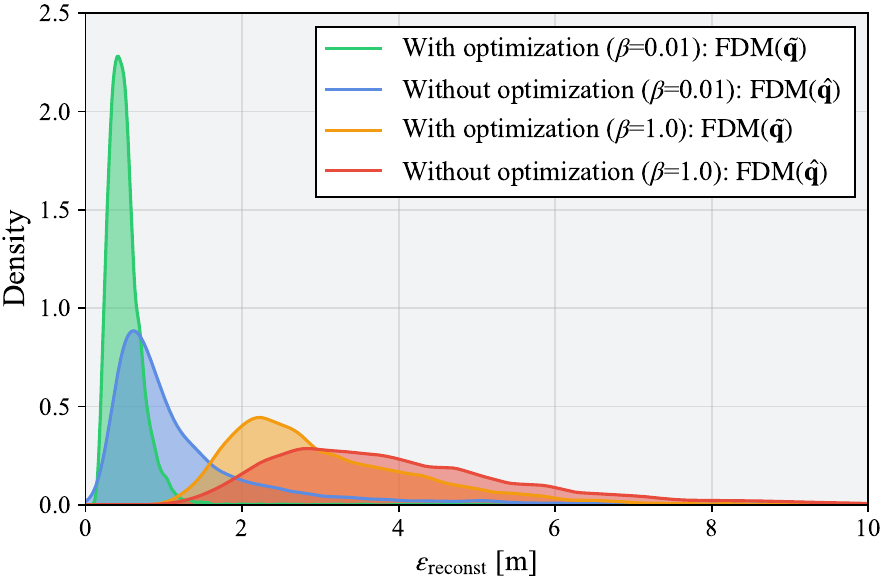}
    \caption{Probability density function of the reconstruction error ($\epsilon_{\mathrm{reconst}}$) for force density-based reconstructions (without optimization) and for reconstructions optimized with a residual force optimization post-processing step. A low KL divergence weight of 0.01 produces better reconstructions than a regular KL divergence weight of 1.0.}
    \label{fig:results_reconst_pdf}
\end{figure}

We find direct geometry predictions (Figure \ref{fig:results_reconstruct}, right) to have a significantly lower $\epsilon_{\mathrm{reconst}}$, though they are not in equilibrium, having a non-zero $\epsilon_{\mathrm{residual}}$. Our combined optimization can typically find an equilibrium geometry close to 
$\hat{\mathbf{x}}$ with minimal changes to $\hat{\mathbf{q}}$ (Figure \ref{fig:results_reconstruct}, bottom). Comparing the pure force density-based reconstruction and the optimization-based reconstruction across the test set we find a consistent reduction in $\epsilon_{\mathrm{reconst}}$ (Figure \ref{fig:results_reconst_pdf}). The optimization reduces the median $\epsilon_{\mathrm{reconst}}$ from 0.90 m to 0.47 m. Due to the long tail in the pre-optimization distribution of $\epsilon_{\mathrm{reconst}}$ the mean is reduced even more drastically, from 2.96 m to 0.61 m.

Figure \ref{fig:results_reconst_pdf} also shows the reconstruction error for a VGAE trained with a KL divergence weight of 1.0. Setting this weight to 1.0 theoretically enables sampling directly from a more strongly regularized latent space, potentially eliminating the need for a diffusion model. In practice, however, we find that increasing the KL divergence weight substantially degrades reconstruction quality before the posterior becomes sufficiently close to the prior for effective sampling.

\begin{figure*}[h]
    \centering
    \includegraphics[width=\linewidth]{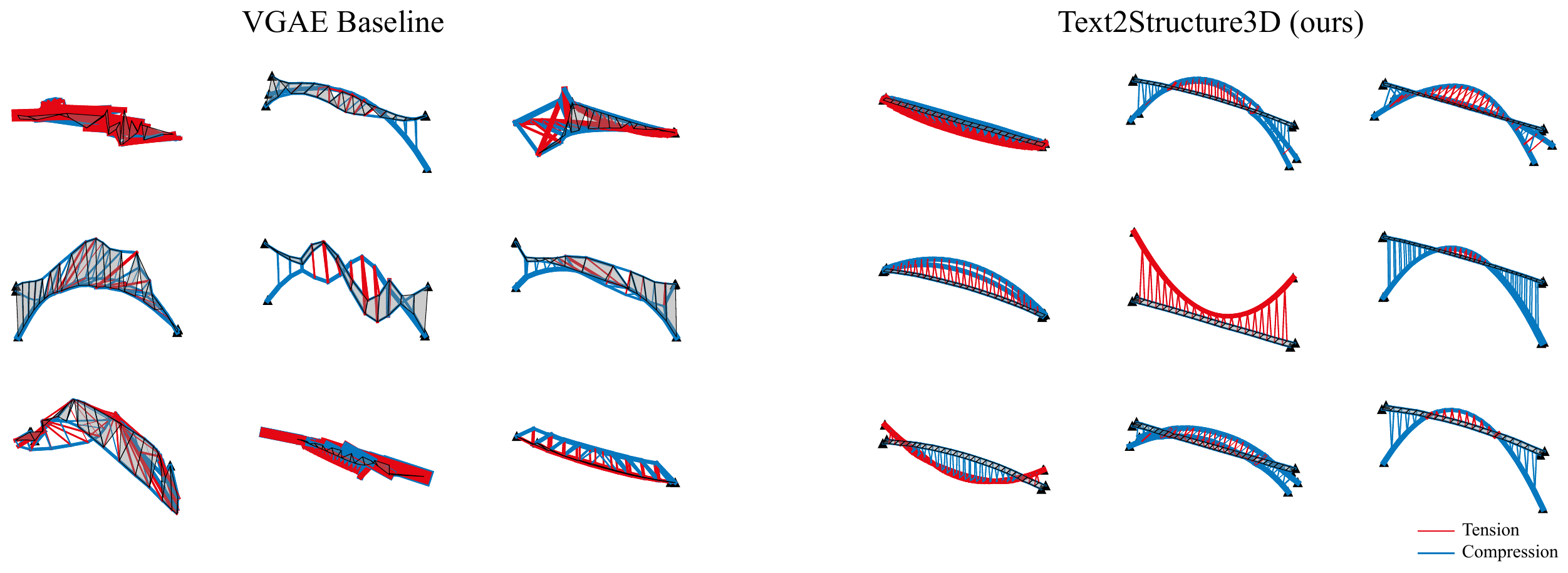}
    \caption{Unconditionally generated equilibrium structures by a VGAE Baseline trained with a KL divergence weight of 1.0 (left) and by Text2Structure3D (right). Line thickness corresponds to force magnitudes in tension (red) and compression (blue).}
    \label{fig:vae_baseline}
\end{figure*}

Figure \ref{fig:vae_baseline} compares random unconditional samples generated by this VGAE baseline by directly sampling from the prior with those generated by Text2Structure3D when no text prompt is provided. The perceptual quality of samples generated using only a VGAE degrades substantially compared to those produced by Text2Structure3D. One way to assess the quality of the generated results quantitatively is through the agreement between the sampled geometry, and the form-found geometry based on the sampled force densities. To this end, we measure the average nodal distance between both samples ($\epsilon_\mathrm{sample}$), similarly to Equation \ref{eq:epsilon_reconst}. When $\epsilon_\mathrm{sample}$ is close to zero, the geometry and force density samples are in agreement, which is a necessary although not sufficient requirement for good sample quality. 

Table \ref{tab:runtime_summary} summarizes the average $\epsilon_\mathrm{sample}$ and generation time across 1000 samples for Text2Structure3D and the VGAE baseline. Text2Structure3D achieves the lowest $\epsilon_\mathrm{sample}$ when optimization is applied (0.40 m), followed by Text2Structure3D without optimization (5.88 m).
Although the VGAE without optimization has by far the shortest generation time (0.01 s), its $\epsilon_\mathrm{sample}$ is much higher (47.40 m), providing a substantially weaker initialization for optimization.
As a result, its total average generation time after optimization still exceeds that of Text2Structure3D with optimization (3.81 s vs. 2.76 s), while not achieving a similar $\epsilon_\mathrm{sample}$ (8.39 m vs 0.40 m).
\ref{sec:time_complexity} provides additional results on the computation time of Text2Structure3D.

\begin{table}[h]
\centering
\resizebox{\columnwidth}{!}{
\begin{tabular}{lcccc}
\hline
 & \multicolumn{2}{c}{VGAE Baseline} & \multicolumn{2}{c}{Text2Structure3D} \\
\hline
Optimization & No & Yes & No & Yes \\
\hline
Generation time [s] $\downarrow$ & \textbf{0.01 $\pm$ 0.00} & 3.81 $\pm$ 2.53 & 0.60 $\pm$ 0.12 & 2.76 $\pm$ 1.29 \\
$\epsilon_\mathrm{sample}$ [m] $\downarrow$ & 47.40 $\pm$ 311.28 & 8.39 $\pm$ 77.7 & 5.88 $\pm$ 15.71 & \textbf{0.40 $\pm$ 2.68} \\
\hline
\end{tabular}}
\caption{Comparison of generation time and sample consistency ($\epsilon_\mathrm{sample}$) for the VGAE baseline and Text2Structure3D, with and without optimization. Values are reported as mean $\pm$ standard deviation over 1000 independent samples, run on a single NVIDIA RTX 6000 Ada GPU.}
\label{tab:runtime_summary}
\end{table}

\subsection{Prompt Adherence}
\label{sec:text2structurerepromptadh}
\begin{figure*}[h]
    \centering
    \includegraphics[width=\linewidth]{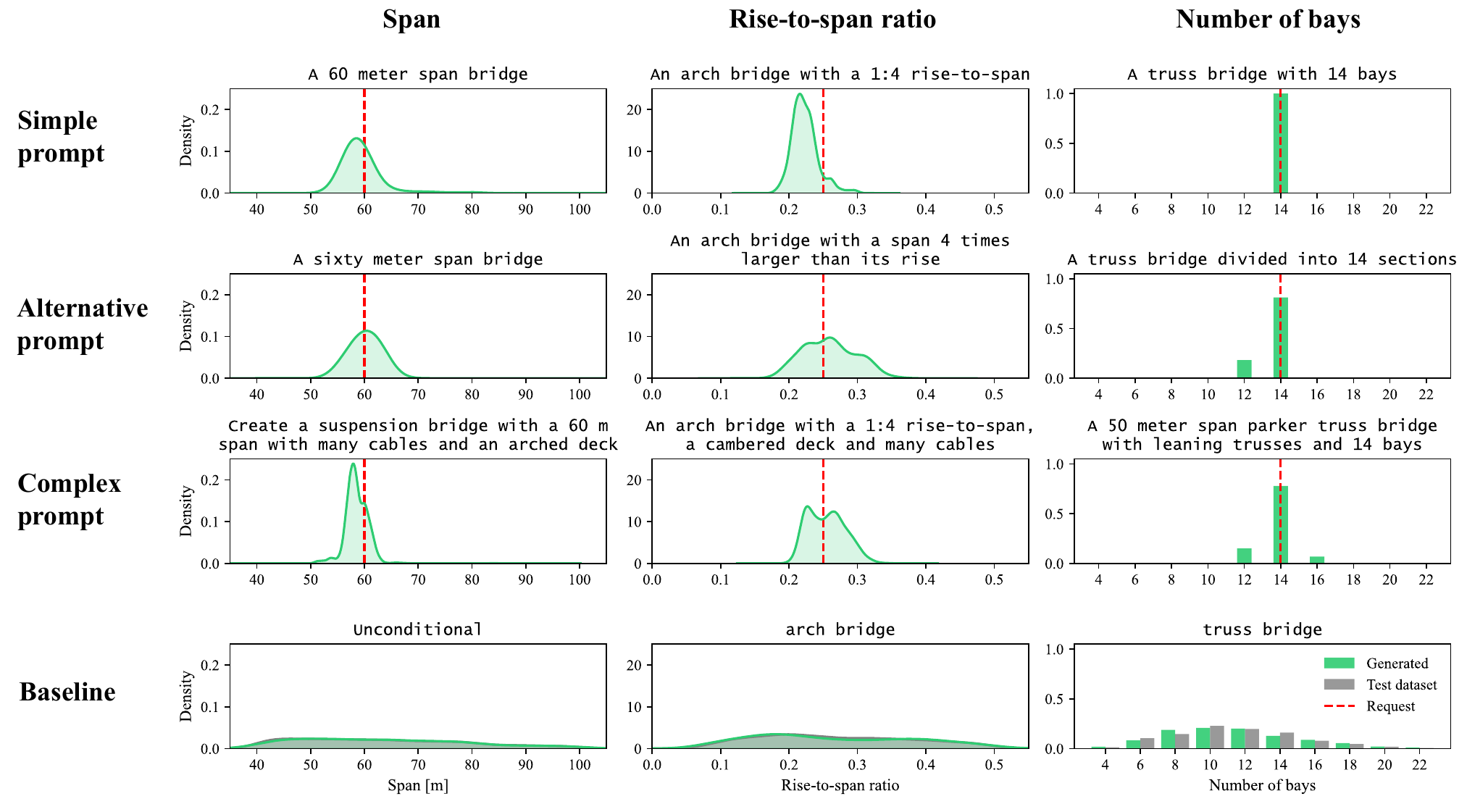}
    \caption{Distributions of generated structures for prompts containing specifications for the span (left), rise-to-span ratio (center) and number of bays (right). For each we evaluate a simple prompt similar to the dataset labels (first row), an alternative prompt worded in a way the model has not encountered during training (second row), and a complex prompt containing multiple requests (third row). As a baseline we evaluate unconditional generation compared to the test dataset distribution (fourth row). For each prompt we show a kernel density estimation or categorical probability distribution derived from 1000 generated samples.}
    \label{fig:text2struct_results_prompt_dists}
\end{figure*}

To evaluate the prompt adherence of Text2Structure3D we study the distributions of measurable attributes of 1000 generated structures for prompts containing specifications for those attributes. For each attribute we evaluate 3 types of prompts: firstly, we use a simple prompt containing a specification for a single attribute, written in a way that is similar to the text labels in the training dataset (Figure \ref{fig:text2struct_results_prompt_dists}, first row). Secondly, we evaluate an alternative prompt with the same information as the first, written in a way the model has not encountered during training (second row). Examples include replacing \textit{60} with \textit{sixty}, or using a longer phrase e.g. \textit{a span 4 times larger than its rise} instead of \textit{1:4 rise-to-span}. Finally we evaluate the performance on more complex prompts that contain specifications for multiple other attributes as well (third row). We compare each prompt against an unconditional baseline and the test dataset distribution (fourth row). For typology-specific attributes, such as rise-to-span ratio, we compare against the distribution of structures only of the relevant typology instead.

Overall we find generated structures to adhere to the prompt, with distributions significantly concentrated around the request. Comparing this to unconditional or typology-only conditional generation we find a much broader distribution of attributes. Distributions are mostly symmetric around the request, with the exception of the simple prompt for the rise-to-span ratio which has a significant bias to the left. In general, we observe a higher variance for continuous attributes than we do for discrete attributes like the number of bays.

The alternative prompt distributions are marginally worse than those of the simple prompt, indicating robustness against the resulting changes in text embeddings obtained from CLIP. Results for complex prompts are also only marginally worse, indicating the model is capable of parsing multiple specifications accurately. Finally, we see that unconditional and typology-only conditional generation tightly matches the distribution of the test dataset.

\begin{figure}[h]
    \centering
    \includegraphics[width=\linewidth]{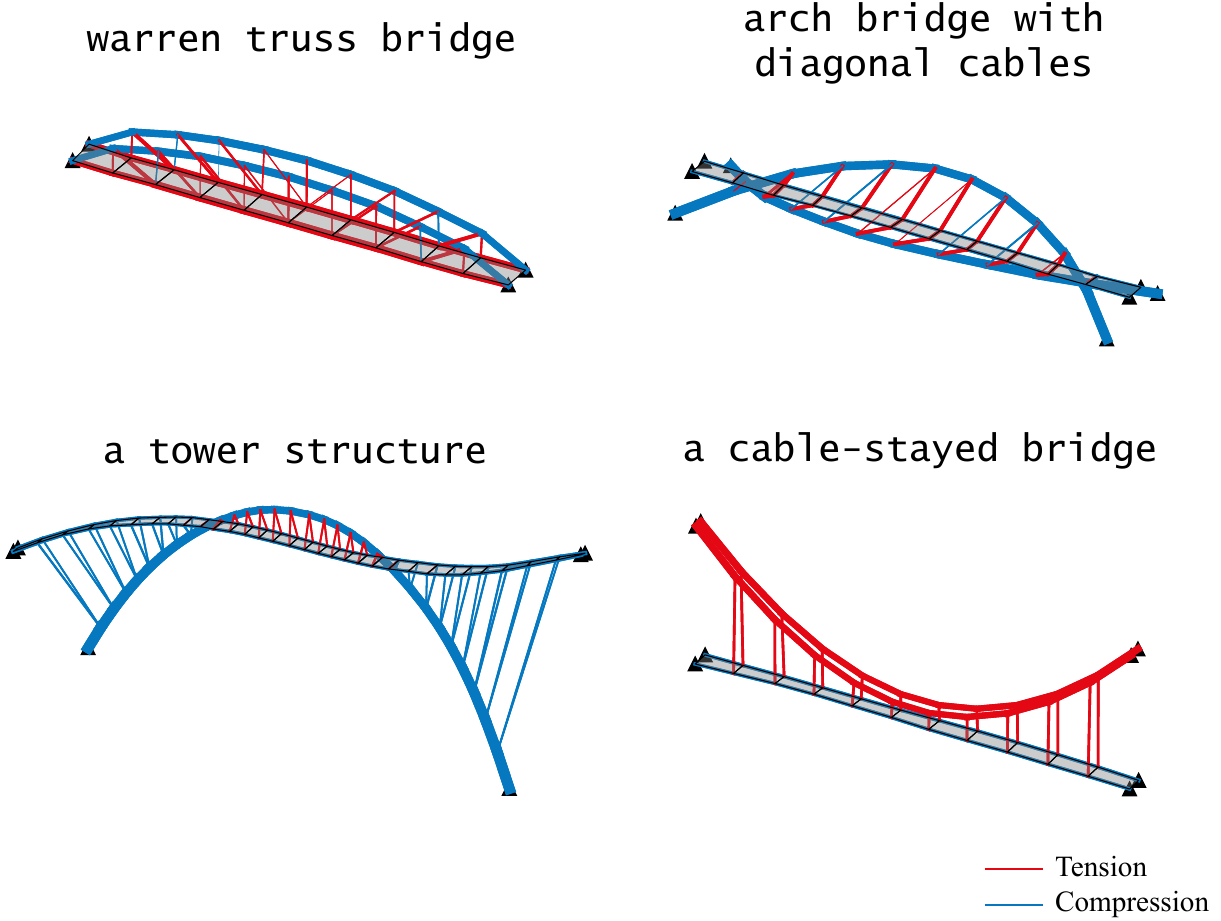}
    \caption{Examples of failure cases. Structures generated from prompts requesting typologies or attributes not present in the dataset generally result in the most semantically similar structures from the training set distribution.}
    \label{fig:failure_cases}
\end{figure}

We furthermore evaluate failure cases in which the model is unable to generate a structure that directly corresponds to the input prompt. Figure \ref{fig:failure_cases} shows generated samples for prompts containing structural typologies or attributes that are not represented in the training dataset. We observe a strong bias toward structures within the training distribution, with the model generally generating the semantically closest known structural configuration. For example, the prompt \textit{warren truss bridge} results in a different truss variant present in the dataset, while the prompt \textit{a cable-stayed bridge} generates a suspension bridge. Similarly, prompts containing unseen attribute combinations, such as \textit{arch bridge with diagonal cables}, are interpreted within the constraints of the learned dataset distribution, in this case producing leaning and intersecting arches. The prompt furthest removed from the training distribution, \textit{a tower structure}, results in a tall arch-like structure. These results further highlight the semantic robustness of the model but also its current dependence on the scope and diversity of the training dataset.

\section{Discussion}
\label{sec:discussion}

\subsection{Limitations}
Text2Structure3D significantly advances the generative modeling of equilibrium structures, yet a number of important limitations remain.

Firstly, while we strived to make our dataset as diverse as possible, we limited the scope of data generation to a reasonable number of bridge typologies. Although Text2Structure3D has been designed with application to more general datasets in mind, several aspects of the model still make use of the limited level of complexity of the dataset presented in this work.

One such limitation is the parametric model-based approach used for the topology MLP. Similar to previous research, this part of Text2Structure3D will face challenges in scalability when applied to more general datasets.

A further limitation concerns the level of design information represented by the current model. Because our dataset only contains bridge structures with a single load case, the loading condition can be derived from the generated geometry. While this is generally sufficient for architectural exploration of structurally plausible forms, it limits the level of control over user-defined load cases, which is especially relevant for structural engineers.

Finally, our model has been trained on structures that are zero-centered with the deck aligned along the x-axis, whereas such a canonical orientation might not be available for more general classes of structures.

\subsection{Practical and Ethical Considerations}
The methods presented in this paper have the potential to support a new design paradigm, and therefore raise a number of practical and ethical considerations.
First and foremost, while these methods can enable architects to design more structurally informed geometries, this does not eliminate the need for engineers in the conceptual design stage. This is particularly true when it comes to engineering responsibility, which should always be with the engineer using a tool rather than with the tool itself.

In this context, the practical role of Text2Structure3D should be understood differently for different design stakeholders. For architects, the method may be particularly useful for exploring structurally informed geometries during early-stage design, especially before an engineer becomes involved.
When structural engineers are involved in early-stage design, it may serve as a source of initial inspiration for alternative equilibrium geometries, especially for engineers with limited experience with structural form-finding.
In all cases, however, the assignment of material properties, cross-sections, and connection parameters, as well as structural analysis, interpretation of results, and subsequent refinement of the generated geometry, remain outside the scope of the current model and fall under the responsibility of the engineer.

Another consideration to be made is the bias of generated designs toward the training dataset. The synthetic dataset presented in this paper is primarily meant to serve as a testbed for generative models such as Text2Structure3D, rather than as a training dataset for a more general model to be used in engineering and design practice. The training dataset for a model intended for real-world use should thus not only be larger and contain more typologies, but should also be sourced in a more democratic way. Creating digital models of existing structures, or crowd-sourcing synthetic data from multiple sources could help avoid strong biases toward individual design assumptions, styles and descriptions.

\subsection{Future Work}
Primary directions for future work include the creation of a broader dataset of equilibrium structures and adapting the generative modeling strategies presented in this work to accommodate such a dataset.

We repeat the common notion in the AEC industry that dataset availability is a key bottleneck to developing ML applications. In the context of this work, a more general dataset of bridge structures containing more typologies such as cable-stayed or cantilever bridges would already make the presented model more powerful. In this regard the model would also benefit from richer and more variously sourced text labels that minimize any particular dataset bias. One way to reduce model bias would be to train on datasets of existing structures, which would also improve the model’s applicability to, and validation against, real-world design cases.

Additional quantitative attributes could be assigned with Finite Element Method (FEM) analyses and further qualitative and subjective attributes could be included using data labeling strategies based on clustering techniques such as Self-Organizing Maps (SOM) \cite{saldanaochoa2021, guo2022}.
We envision the inclusion of such additional quantitative attributes, combined with a more general ability to specify load cases and constraints, would significantly enhance the applicability of the proposed model.

Beyond the information included in the text labels, more varied wording could be achieved using LLMs to further increase the robustness of the model. An additional use of LLMs could be the implementation of Text2Structure3D in an agentic framework, expanding the ways in which users can interact with the model. Furthermore, another opportunity for enhancing user interactions with the model is to include explainable AI techniques, though also this is beyond the scope of the current work.

Development of a more general dataset containing typologies beyond bridge structures would need to be accompanied by advances in a few key areas of the generative modeling strategy as well. Most notably, potential alternatives for the parametric model-based strategy for predicting the topology include ground structure-based approaches, similar to link prediction applications, or autoregressive approaches, for example using reinforcement learning. To overcome the challenges of training on datasets without a clear canonical orientation, Equivariant Graph Neural Networks (EGNN) could provide a solution \cite{bleker2025generalized}.

To maximize the practical impact of generative models it is necessary to integrate them into existing user workflows. For architectural practice the most relevant integration would be in Building Information Modeling (BIM) and Computer-Aided Design (CAD) environments, such as Rhinoceros and Grasshopper. Moreover, engineers could also benefit from the integration of Text2Structure3D or similar models directly in existing FEM software.

Finally, while in this work we focused on text-conditioning, the presented model architecture is compatible with other types of modalities for conditioning as well.
Most prominently, this opens the opportunity for generating structures from images, such as design sketches, screenshots of CAD models or even photos of existing structures.

\section{Conclusion}
\label{sec:conclusion}
We presented Text2Structure3D, a graph-based generative model for equilibrium structures based on latent diffusion conditioned on natural language.
Text2Structure3D allows the generation of graph-based structures of various topology, geometry and force configuration, greatly improving the generalization abilities of existing generative approaches built around parametric models in structural design.
To train Text2Structure3D, we developed a cross-typological dataset of bridge structures combined with text descriptions as labels.
Generated structures show strong adherence to the specifications in the text prompt, even when provided with prompts containing multiple attribute specifications or alternative wording to the prompts in the training dataset.
To adapt latent diffusion for equilibrium structures, we introduce a residual force optimization step that can guarantee predicted structures are in equilibrium while improving reconstruction performance.
Current limitations include the limited scope of the training dataset and the parametric model-based approach for topology predictions.
Looking ahead, by addressing these limitations Text2Structure3D could have the potential of forming the basis for a general-purpose foundation model for structural design applications.
\FloatBarrier

\section*{CRediT authorship contribution statement}
\textbf{Lazlo Bleker}: Conceptualization, Data Curation, Formal Analysis, Investigation, Methodology, Software, Validation, Visualization, Writing -- original draft, Writing -- review \& editing.
\textbf{Zifeng Guo}: Conceptualization, Investigation, Methodology, Software, Validation, Writing -- review \& editing.
\textbf{Kaleb Smith}: Conceptualization, Methodology, Validation, Resources, Writing -- review \& editing.
\textbf{Kam-Ming Mark Tam}: Conceptualization, Methodology, Supervision, Validation, Writing -- review \& editing.
\textbf{Karla Saldaña Ochoa}: Conceptualization, Funding acquisition, Methodology, Supervision, Validation, Resources, Writing -- review \& editing.
\textbf{Pierluigi D’Acunto}: Conceptualization, Funding acquisition, Methodology, Supervision, Validation, Resources, Writing -- review \& editing.

\section*{Declaration of Competing Interest}
The authors declare that they have no known competing financial interests or personal relationships that could have appeared to influence the work reported in this paper.

\section*{Declaration of Generative AI and AI-assisted technologies in the writing process}
During the preparation of this work the author(s) used large language models in order to improve sentence structure. After using this tool/service, the author(s) reviewed and edited the content as needed and take(s) full responsibility for the content of the publication.

\section*{Data Availability}
Data will be made available upon request.

\section*{Acknowledgments}
This work is supported in part by the International Graduate School of Science and Engineering (IGSSE) of the Technical University of Munich (TUM). The authors gratefully acknowledge the NVIDIA AI Technology Center at the University of Florida for its technical support and for granting access to computing resources.

\appendix
\section{Implementation Details}
\label{sec:apdx_implementation_details}

\begin{figure*}[h]
    \centering
    \includegraphics[width=\linewidth]{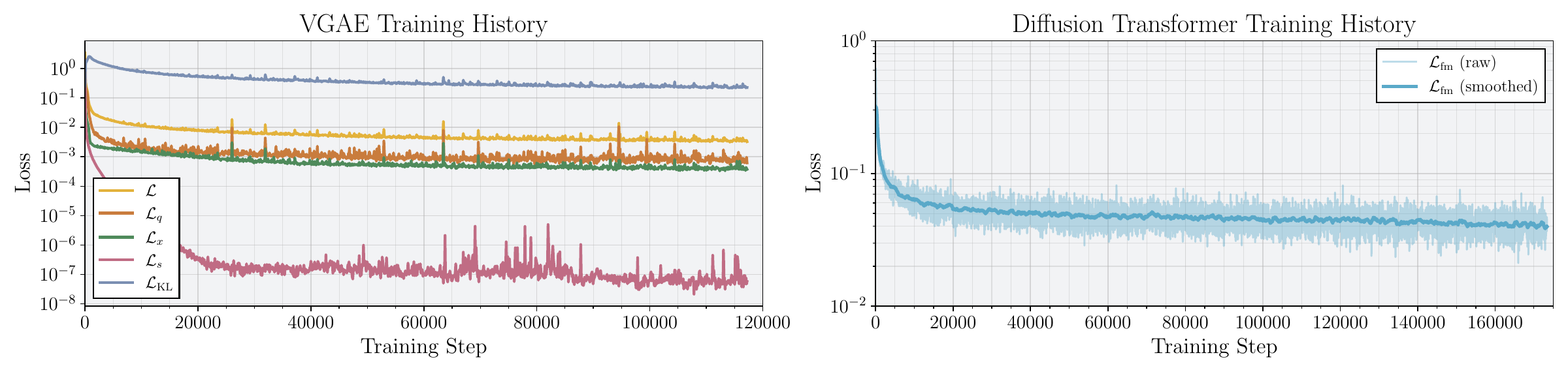}
    \caption{Training history for the VGAE (left) and Diffusion Transformer (right). Both models are trained with a fixed time budget of 72 hours. Shown are the total VGAE loss ($\mathcal{L}$), the reconstruction loss for force densities ($\mathcal{L}_q$), coordinates ($\mathcal{L}_x$), supports ($\mathcal{L}_s$), the KL divergence loss ($\mathcal{L}_{\text{KL}}$) and the flow-matching loss ($\mathcal{L}_{\text{fm}}$).}
    \label{fig:training_history}
\end{figure*}

\textbf{Variational Graph Auto-Encoder.} The VGAE \cite{kipf2016} decoder and encoder are based on a graph transformer as described by Rampá\v{s}ek et al. \cite{rampasek2022}. For the message passing layer we use a GINE convolution \cite{hu2020}, and for global attention a standard Transformer encoder block \cite{vaswani2017} with 4 attention heads. We set the latent dimension $d_z=8$ and use 9 layers with hidden dimension $d_h=128$. We train the VGAE with a fixed time budget of 72 hours on a single NVIDIA RTX 6000 Ada GPU with Adam \cite{kingma2015} using a learning rate of $1e-4$ and a batch size $b=128$. Figure \ref{fig:training_history} (left) shows the training history of the total VGAE objective and its constituent weighted loss terms.

\textbf{Diffusion Transformer} We use a largely unmodified DiT architecture \cite{peebles2023} using 12 layers with a hidden dimension $d_h=1536$ and 24 attention heads. We use both general graph positional and structural encodings, as well as positional encodings specific to our dataset. For general encodings we apply Laplacian Positional Encodings (LPE) \cite{dwivedi2023} using 12 eigenvectors and Random-Walk Structural Encodings (RWSE) \cite{dwivedi2022} using 12 random walks. For bridge-specific encodings we add integer indicators for nodes being part of the deck or not, on which side of the bridge they belong to—both along and orthogonal to the span—and the shortest n-hop distance to one of the supports. We train the DiT on the same hardware and 72-hour time budget as the VGAE with Adam using a learning rate of $1e-4$ and a batch size $b=64$. Figure \ref{fig:training_history} (right) shows the training history for the flow-matching loss.

\textbf{Topology MLP.} The Topology MLP consists of 2 layers with a hidden dimension $d_h=512$ followed by a ReLU activation function \cite{nair2010} and a dropout rate of $0.1$. We use Adam with a learning rate of $1e-3$ with a batch size of $b=128$ and train the MLP for 20 minutes on an NVIDIA RTX 4080 which we found to be sufficient for the model to converge.

In general, the chosen hyperparameters for all models are a result of exploratory tuning, balancing performance and computational cost rather than exhaustive optimization. Further performance improvements may be achievable through more extensive hyperparameter optimization.

\section{Time Complexity}
\label{sec:time_complexity}

\begin{figure}[h]
    \centering
    \includegraphics[width=\linewidth]{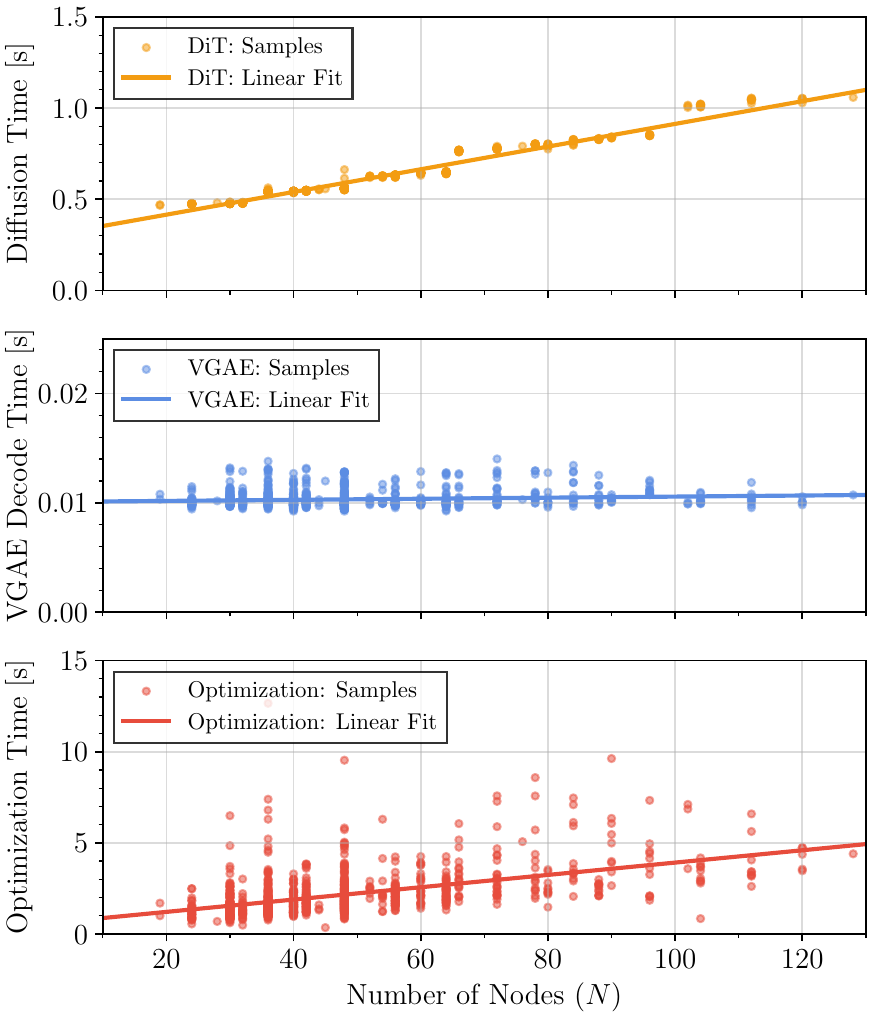}
    \caption{Computation time vs. number of nodes for the Diffusion Transformer (DiT) sampling process (top), Variational Graph Autoencoder (VGAE) decoder (center), and residual force optimization (bottom). Shown are 1000 independent samples and their linear fit. Reported inference times are for a single NVIDIA RTX 6000 Ada GPU.
    \label{fig:time_complexity}}
\end{figure}

To further evaluate the computational performance of Text2Structure3D, we analyze the generation time of its main inference components as a function of the number of nodes $N$ in the generated structural graph. Figure \ref{fig:time_complexity} reports the measured runtime for 1000 generated samples, together with a linear fit for each component. The analysis separates the runtime of the Diffusion Transformer (DiT), the VGAE decoder, and the residual force optimization step.

The DiT runtime increases approximately linearly with the number of nodes over the evaluated range. This trend is partly expected, as diffusion is performed on a node-level latent representation and therefore requires processing a larger set of graph tokens as $N$ increases. However, because the DiT uses an attention-based architecture, its self-attention operations are expected to have a quadratic complexity in the number of nodes. The close-to-linear trend observed here might therefore only reflect the limited graph-size range considered in this study and may not hold for substantially larger structures. The VGAE decoder is the fastest component and remains nearly constant across the evaluated range, indicating that its runtime is largely dominated by fixed overhead. In contrast, the residual force optimization step shows the largest variance and represents the dominant source of computational cost for most graphs. Although its average runtime also follows an approximately linear trend over the tested range, individual samples can require substantially longer optimization times depending on the quality of the initialization and the difficulty of satisfying the equilibrium constraints.

\bibliographystyle{elsarticle-num}
\bibliography{refs}

\end{document}